\documentclass[prd,aps,floats,twocolumn]{revtex4}

\usepackage{amssymb}
\usepackage{amsmath}
\usepackage{graphics}

\input epsf
\def\ba{\begin{eqnarray}}
\def\ea{\end{eqnarray}}
\def\be{\begin{equation}}
\def\ee{\end{equation}}

\begin{document}

\title{String Propagation through a \\
Big Crunch/Big Bang Transition}

\author{Andrew J. Tolley$^1$}

\affiliation{$^1$Joseph Henry Laboratories,
Princeton University,\\
Princeton, NJ 08544, USA}

\begin{abstract}
We consider the propagation of classical and quantum strings on cosmological space-times which interpolate from a collapsing phase to an expanding phase. We begin by considering the classical propagation of strings on space-times with isotropic and anisotropic cosmological singularities. We find that cosmological singularities fall into two classes, in the first class the string evolution is well behaved all the way up to the singularity, whilst in the second class it becomes ill-defined. Then assuming the singularities are regulated by string scale corrections, we consider the implications of the propagation through a `bounce'. It is known that as we evolve through a bounce, quantum strings will become excited giving rise to `particle transmutation'. We reconsider this effect, giving qualitative arguments for the amount of excitation for each class. We find that strings whose physical wavelength at the bounce is less that $\sqrt{\alpha'}$ inevitably emerge in highly excited states, and that  in this regime there is an interesting correspondence between strings on anisotropic cosmological space-times and plane waves. We argue that long wavelength modes, such as those describing cosmological perturbations, will also emerge in mildly excited string scale mass states. Finally we discuss the relevance of this to the propagation of cosmological perturbations in models such as the ekpyrotic/cyclic universe. 

\end{abstract}

PACS number(s): 11.25.-w,04.50.+h, 98.80.-k

\maketitle

\section{Introduction}

Recent ideas have brought about a revival of interest in the idea
that the big bang may not be the beginning of time, and that the
universe could undergo a transition from a collapsing phase to an
expanding phase (referred to as a `bounce'). In particular, the ekpyrotic  \cite{ekpyrotic} and cyclic \cite{cyclic} models
which make use of this idea have been put forward as possible
alternatives to standard inflationary cosmology. The cyclic model
reinterprets the conventional big bang as a collision of two
orbifold planes in heterotic M-theory \cite{heterotic}. A natural question to ask in any such model is what happens to cosmological perturbations as we evolve through the bounce (see \cite{perts} for various approaches to this problem)? If the bounce occurs close to the string scale, it is important to understand how string scale physics effects the propagation of cosmological perturbations \cite{porrati}. This is similar to the trans-Planckian problem in inflation \cite{transplanck}. In the ekyrotic/cyclic context, scalar and tensor perturbations are coherent states of dilatons and gravitons respectively. Since the dilaton and graviton are just excitation states of a quantum string, we can gain insight into what happens for cosmological perturbations by looking directly at how classical and quantum strings evolve through the bounce. This allows us to understand the cosmological implications of the string scale physics. 

In the following we shall focus on bounces that occur at weak string coupling where perturbative string worldsheet methods are appropriate. Cosmological singularities at weak coupling remain the most promising to understand since the formalism to describe them exists, whereas far less is known about strings at strong coupling in the time dependent context. Unfortunately little direct progress has been made on these questions since the string geometry becomes highly curved and the usual $\alpha'$ expansion of string theory breaks down. Thus it is technically far too difficult to compute the renormalization group (RG) equations necessary to check the conformal invariance of the string sigma model. 

In this paper we take a more modest approach in order to gain insight into these questions. We assume the existence of weak coupling bouncing solutions to the string RG equations and consider the propagation of strings in the background of these solutions. In this sense we are using the strings as probes of the geometry. Although we cannot rigorously quantize the string in these backgrounds, we can  anticipate some of the essential physics by making a useful analogy with strings in plane wave space-times and by a semiclassical treatment. There has been considerable earlier work on the properties of strings on cosmological space-times (for a review of some of this work see \cite{book,review}). In particular ref. \cite{Horowitz,Sanchez2} considered the implications of string propagation on a big crunch/big bang space-time in the context of null cosmologies or plane waves where the string sigma model may be solved exactly. The main conclusion of these works are that the strings may emerge from a bounce in highly excited states. The space-time interpretation of this phenomena is `particle transmutation', in which particles change their mass and spin. We may think of this as arising from the interaction of incoming particles with the coherent state of gravitons that make up the background geometry. As a result a gravitational wave in the collapsing phase may emerge as a field configuration of different spin with string scale mass. There has also been considerable earlier work on the behavior of strings on cosmological space-times focusing principally on expanding ones \cite{Sanchez1}  where it has been observed that the string modes freeze, i.e. stop oscillating \cite{veneziano2,nullstrings,review}, which is analogous to the freezing of cosmological perturbations in inflation or ekpyrosis as their wavelengths cross the Hubble horizon. More recent work on strings on cosmological space-times has focused on the notion of time-dependent orbifolds \cite{orbifolds,Seiberg,Seiberg2,HorowitzPolchinski}, whereby regular manifolds are identified under boosts or combinations of boosts, translations and rotations. In the simplest examples these geometries are regular everywhere except at special points or subspaces where the manifold is non-Haussdorf. These constructions often introduce additional complications such as closed time-like curves, additional non cosmological regions and their validity as consistent string backgrounds has been questioned \cite{Seiberg,HorowitzPolchinski,orbifoldreview}. It is not clear however what the study of these special cases implies for the more generic case where the space-time curvature blows up near the singularity and string $\alpha'$ effects become important. Already from the point of view of low energy supergravities, these space-times are typically unstable with respect to the formation of Kasner-like curvature singularities. In the following it is these type of singularities that we shall consider. 

We argue that bouncing solutions fall into two classes from the perspective of string propagation depending on the nature of the singularity that the bounce regulates. For isotropic space-times, if the singularity occurs at finite conformal time, then we find that the evolution of strings is only mildly sensitive to what happens at the bounce. If however the singularity occurs at infinite conformal time, then the period of the bounce will be a very long period in terms of conformal time and the evolution of the strings will consequently be more sensitive to the bounce.  

As the universe collapses, strings describing short wavelength modes effectively become tensionless and begin to move along null geodesics. By short wavelength we mean modes whose physical wavelength is much less than the string scale at the bounce. These are the string frame analogue of trans-Planckian modes \cite{transplanck}, except that since $g_s \ll 1$, we use the string scale as the relevant cutoff. Due to the gravitational blueshift and subsequent redshift, these may correspond to observable length scales today. In this regime, the evolution of the string becomes nearly identical to the evolution of a string on a plane wave geometry which is constructed by taking the Penrose limit of the cosmological geometry in the direction that contracts most rapidly.  Since the latter are exactly solvable conformal field theories we may use aspects of this correspondence to anticipate the properties of quantum strings on cosmological space-times. In fact this is closely connected to the null-string expansion \cite{nullstrings} which has been used to study strings on cosmological space-times. The fact that strings become tensionless near the singularity was used in \cite{nullstrings} and more recently in \cite{TPS} where it was used to argue that near the singularity, strings may be well described by a $1/\alpha'$ expansion. 

By contrast strings describing long wavelength modes (cis-Planckian) may no longer be thought of as tensionless and have a very different behavior. Here we find that the modes freeze and string excitation only occurs when the space-time curvature becomes comparable to the string scale which we assume to be close to the bounce. The amount of excitation is independent of wavelength, and so depending on precisely what happens in the string scale regime, modes of arbitrarily long wavelength may emerge from the bounce in highly excited states. 

One motivation for this work is the proposal of ref. \cite{TPS} to continue strings and membranes through the collision of two orbifold planes, which is relevant to the ekpyrotic and cyclic universe models. Here we shall view things from the 10d string perspective where we see that classical strings may be continued through the singularity. We anticipate that short wavelength modes will emerge from the singularity in highly excited states. Cosmologically these modes are not a problem and they may have interesting consequences such as the formation of small black holes. However, semi-classical arguments indicate that if the bounce occurs close to the string scale, even strings describing long wavelength super-horizon modes will emerge from the bounce in excited states, having potentially important implications for the propagation of cosmological perturbations in the ekpyrotic/cyclic models. Although these stringy states are short lived because of $g_s$ effects, they will decay back into massless and low mass perturbations and have an indirect effect on the late time cosmological power spectrum. Since this effect occurs close to the string scale, semiclassical arguments are not appropriate, nevertheless we anticipate a similar behavior in a full quantum theory.

We begin in section \ref{classicalstrings1} with a discussion of classical strings on cosmological spacetimes and discuss the conditions for the string modes to freeze near the singularity. In section \ref{spacelike} we elucidate the connection between strings describing short wavelength modes and plane waves and in section \ref{nullcosmological} we discuss the quantum aspect of string mode excitation. We conclude in section \ref{ekpyrotic/cyclic} by connecting these ideas back to the ekpyrotic and cyclic models. 
 
\section{Classical strings on cosmological space-times}

\label{classicalstrings1}

In this section we consider the propagation of classical strings on isotropic and anisotropic spatially flat cosmologies. The restriction to spatially flat geometries is for simplicity and we expect similar qualitative conclusions in the general case. For the types of bouncing geometries we consider, the solutions of the leading order in $\alpha'$ string equations of motion give rise to cosmological singularities. We have in mind that the singularities are regulated by $\alpha'$ effects or similar string scale physics. We can learn how sensitive the propagation of strings is to the regulating physics of the bounce by looking at the evolution of the string towards the cosmological singularity. We find that the behavior splits into two classes. In class I the evolution of strings is well behaved all the way to the singularity. In these cases we anticipate that the strings will only be mildly sensitive to the physics of the bounce. For class II space-times the evolution of the strings becomes ill-defined at the singularity. In these cases we anticipate a strong sensitivity to the physics of the bounce. In order to make this behavior transparent, we formulate the equations of motion for a string as a Hamiltonian system. For class I space-times, the position and momentum of the string $(x^i(\sigma),P_i(\sigma))$ are finite at the cosmological singularity. For class II space-times they become ill-defined since $x^i(\sigma)$ becomes infinite. The behavior of classical strings has also been considered recently in ref. \cite{Gustavo} which takes a more detailed look at explicitly matching string solutions across a singularity. 

In the rest of the paper we shall use several different time variables, our conventions are $t$ for conformal time, $\tau$ for worldsheet time, and $x^0$ for proper time, except where otherwise stated. The spacelike worldsheet coordinate is denoted by $\sigma$.

\subsection{Isotropic case}

First let us consider the isotropic case. Our
approach shall closely follow that of ref. \cite{TPS}. We
begin with the classical Nambu-Goto action for a string on a fully
isotropic FRW space-time in conformal time coordinates $x^{\mu}=(t,\vec{x})$, $g_{\mu \nu}=a(t)^{2}\eta _{\mu \nu}$, 
\ba \nonumber S&=&-\mu\int d^{2}\sigma
\sqrt{-\det \left( a(t)^{2}\eta _{\mu \nu }\partial _{a}x^{\mu
}\partial _{b}x^{\nu } \right) } \\
&=&-\mu\int d^{2}\sigma a^{2}(t)
\sqrt{-\det \left ( \eta _{\mu \nu }\partial _{a}x^{\mu }\partial
_{b}x^{\nu } \right )}, 
\ea 
where $\mu=1/(2\pi\alpha')$ is the
string tension, $d^2 \sigma=d\tau d\sigma$, and for a closed string $0<\sigma<2\pi$. This is analogous to a string on Minkowski space-time
with variable tension $\mu(t)=\mu a^2(t)$. We may use the
worldsheet diffeomorphism invariance to choose the `background time' gauge (worldsheet = target space conformal time) $t=\tau $. We may further choose $\chi=\dot{\vec{x}} \cdot
{\vec{x}}' = 0$. Here $\dot{A}=A_{,\tau}$ and $A'=A_{,\sigma}$. To prove this we choose it to vanish at a given
time and then one can show that the equations of motion imply $\dot{\chi}=0$. This follows an approach taken in  ref. \cite{cosmic}. 
The action is then 
\be S=-\mu\int
dtd\sigma a^{2}(t)\left| {\vec{x}}^{\prime }\right|
\sqrt{1-|\dot{\vec{x}}|^{2}}, \ee and the momentum conjugate to
$\vec{x}$ is,
\be \vec{P}(t,\sigma)=\frac{\mu a^{2}(t)\left|
{\vec{x}}^{\prime }\right| \dot{\vec{x}}}{\sqrt{1-|
{\dot{\vec{x}}}|^{2}}} \, . 
\ee
Defining the Hamiltonian in the
usual way 
\ba H &=& \int d\sigma \left[ \vec{P} \cdot
\dot{\vec{x}}+\mu a^{2}(t) |{\vec{x}}^{\prime}|
\sqrt{1-|\dot{\vec{x}}|^{2}} \right]  \\
&=& \int d\sigma \sqrt{\vec{P
}^{2}+\mu^{2}a^{4}(t) |\vec{x}^{\prime}| ^{2}}, 
\ea 
we obtain
the canonical equations of motion
\ba 
\dot{\vec{x}}&=&\frac{\vec{P}}{\sqrt{\vec{P
}^{2}+\mu^{2}a^{4}(t)|\vec{x}^{\prime }| ^{2}}}, \\
\dot{\vec{P}}&=&{\frac{\partial}{\partial \sigma}}\left(
\frac{\mu^{2}a^{4}(t){\vec{x}}^{\prime }}{ \sqrt{\vec{P
}^{2}+\mu^{2}a^{4}(t)|\vec{x}^{\prime
  }|^{2}}}\right)
\, . 
\ea It is clear that for any FRW space-time for which the
singularity occurs at finite conformal time, then as $a\rightarrow
0$ these equations are well defined and asymptotically we have\be
\dot {\vec{x}}=\frac{\vec{P}}{\sqrt{\vec{P }^{2}}},
\qquad \dot{\vec{P}}=0, \qquad \rightarrow \qquad |\dot{\vec{x}}|=1,
\ee corresponding to a string moving at the speed of light. The
requirement that the singularity occur at finite conformal time is
crucial since the solution asymptotes to \be
\vec{x} \rightarrow \vec{x}_0+\frac{\vec{P_0}}{\sqrt{\vec{P_0}^{2}}} t, \ee and
so if $t$ tends to infinity $\vec{x}$ would diverge. However, no
other assumption about the behavior of $a(t)$ is necessary.

As an example solution, suppose we compactify on a $T^{8}$ so that for $%
i=2...9$ we identify $x_{i}\sim x_{i}+2\pi \theta _{i}$. Then a
closed string that winds in a time-independent way around the eight
toroidal directions has 
\be x_{i}=x_{i}^{0}+n_{i}\theta _{i}\sigma
, \ee 
where $0 \le \sigma < 2\pi$ and $n_{i}$ is the winding number.
The constraint $\dot{\vec{x}} \cdot {\vec{x}}^{\prime}=0$ implies that $x_{1}^{\prime }=0$ and so the
Hamiltonian is 
\be 
H=\sqrt{ \vec{P
}^{2}+\mu^{2}a^{4}(t)|\vec{x}^{\prime }|^{2}}=
\sqrt{P_{1}^{2}+\mu^{2}a^{4}(t) ( \sum_{i}n_{i}\theta
_{i} )^{2}}, 
\ee 
and the only non-trivial equations of motion
are \be \dot{x_{1}}=\frac{P _{1}}{\sqrt{P
_{1}^{2}+\mu^{2}a^{4}(t)\left( \sum_{i} n_{i}\theta _{i}\right)
^{2}}}, \qquad \dot{P_{1}}=0 .\ee Since $P_1$ is
conserved the solution is formally \be
x_{1}(t)=x_1(t_0)+\int_{t_{0}}^{t}dt\frac{1}{\sqrt{1+ \mu^{2}a^{4}(t)
\left( \sum_{i}n_{i}\theta _{i}\right) ^{2}/P _{1}^{2} } } ,\ee which
is a manifestly finite integral for any scale factor $a(t)$ showing
that the classical string behavior for these modes winding around
the $T^{8}$ is finite.

\subsection{Classical strings on anisotropic space-times}

We have seen that on a fully isotropic space-time, the evolution of
strings remains well defined at a space-like singularity. Let us now
see how this situation changes on the more generic case of an
anisotropic singularity. We shall assume that one of the spatial
directions contracts more rapidly than the others \be
ds^{2}=a_1^{2}(t)(-dt^{2}+d{x_1}^{2})+\sum_{i=2}^{9}
a_{i}^{2}(t)dx_{i }^{2}. \ee We choose the parametrization
so that the $a_1(t)$ contracts faster or the same as $a_{i}(t)$
as we go towards the singularity, so that
$r_{i}(t)=a_{i}(t)/a_1(t)$ tends to infinity or a
nonzero constant (note that some of the directions may be expanding). Similarly it is useful to order the $a_{i}$'s so that $a_{i}$ contracts more rapidly or at the same rate as $a_{i+1}$. Then following the same procedure as before in the gauge
$t=\tau $ and $\chi=a_1^{2}(t) \dot{x} \cdot
{x}^{\prime }+\sum_{i} a_{i}^{2}(t) \dot{y_{i}} y_{i}^{\prime }=0$, 
which again may be
justified by showing that $\dot{\chi}=0$, the action is given by 
\be
S= 
-\int dtd\sigma \lambda^2/\epsilon, 
\ee
where we have defined 
\ba
\lambda
&=&\mu a^{2}(t)\sqrt{\left| {x}^{\prime }\right|
^{2}+\sum_{i}r_{i}^{2}(t)\left| x_{i}^{\prime
}\right| ^{2}}, \\
\epsilon &=&\lambda /\sqrt{1-|
\dot{x_1}|^{2}-\sum_{i} r_{i}^{2}(t)|\dot{
x_{i}}|^{2}}.
\ea
We find the conjugate momenta \be
P_1=\epsilon \dot{x}, \qquad  P _{i}=\epsilon r_{i}^{2}\dot{x_{i}}, \ee and the
Hamiltonian is given by
 \ba
 \begin{array}{l}
 H= \int d\sigma \epsilon \\
 \nonumber
 =\int d\sigma \sqrt{P_1^{2}+\mu^{2}a_1^{4}(t)
\left| x_1^{\prime }\right| ^{2}+\sum_{i=2}^9 \left(
 \frac{1}{r_{i}^{2}}P _{i}^{2}+\mu^{2}a_1^{2}a_{i}^{2}\left|x_{i}^{\prime }\right|^{2}\right) } .
\end{array}
\ea
Clearly if $r_{i}(t)$
tends to a nonzero constant (as in the isotropic case) as $a_1(t) \rightarrow 0$ then this Hamiltonian is
well defined as above for arbitrary $a_1(t)$ and so the classical
string solutions on this space-time will be well defined through the
crunch. The qualitative behavior near the singularity may now be
split into 3 classes depending on the nature of the anisotropic collapse:
\begin{eqnarray} a_1^2 a_i^2 \rightarrow 0 \ &&\text{ for }  i=2, \cdots, I \\
a_1^2 a_j^2 \rightarrow c_i={\rm constant} \ &&\text{ for }  j=I+1, \cdots, J \\
a_1^2 a_k^2 \rightarrow \infty \ &&\text{ for }  k=J+1, \cdots, 9
\end{eqnarray} 

{\bf{I.}} If $I=9$ then we find near
the singularity assuming that $a_1$ contracts more rapidly than $a_2$ \be H\rightarrow \int d\sigma |P_1|,
\ee with asymptotic equations of motion \be \dot{x_1}=\frac{P_1}{|P_1|}, \qquad
\dot{x_{i
}}=\frac{P_{i}}{r_{i}^2|P_1|}\rightarrow
0 \qquad \dot{P_1}=\dot{P _{i
}}=0. \ee The string stops oscillating in the directions that
contract less rapidly, and temporarily moves at the speed of light
in the directions that contract most rapidly. Thus it is the
directions that blueshift most rapidly that dominate the behavior
near the singularity. More explicitly we find near $t=0$ \ba
\nonumber x^1&=&x^1_0+\frac{P_0}{|P_0|}
t, \\
x^{i} &=& x^{i}_0+\frac{P_0^{i}}{|P_0|}
\int \frac{1}{r_{i}^2} dt.
\ea

{\bf{II.}} If $I<9$ and $J=9$ then we find a similar behavior except now
there is an effective mass and so the string moves at less than the
speed of light in the $\vec{x}$ directions, as is evident in the
equations of motion \ba
\dot{x_1}&=&\frac{P_1}{\sqrt{P_1^{2}+\sum_{j=I+1}^{J} \mu^{2}c_{j}( \left|
x_{j}^{\prime }\right|^{2}) }} \\
\dot{x_{i}}&=&\frac{P_{i}}{r_{i}^2\sqrt{P_1^{2}+\sum_{j=I+1}^J \mu^{2}c_{j}( \left|
x_{j}^{\prime }\right|^{2}) }}. \ea

{\bf{III.}} If $J<9$,
for at least one direction $x_k$ then the Hamiltonian is well
defined only for the modes for which the associated $x_{k}^{\prime }=0$. However for modes for which the strings oscillate in
the $x_{k}$ directions, the Hamiltonian is dominated by the
direction that expands most rapidly 
\ba 
\nonumber
H \rightarrow \int d\sigma \sqrt{%
\mu^{2}a_1^{2}(t)\left( \sum_{k=J}^{9} a_{k}^{2}\left| x_{k}^{\prime
}\right| ^{2}\right) }, 
\ea 
the Hamiltonian consequently diverges and
so the evolution of these modes is ill-defined at the singularity.

\subsection{Kasner versus 2d Milne space-time}

Let us now see how these conditions relate to some familiar anisotropic cosmological spacetimes. All the low-energy string theory actions contain the
Einstein-dilaton term \be S=\frac{1}{2\kappa_{10}^2}\int d^{10} x
\sqrt{-g} e^{-2\Phi}(R+4\partial_{\mu} \Phi
\partial^{\mu} \Phi), \ee
where the closed string coupling is $g_s=e^{\Phi}$ and we have written the metric in string frame. For simplicity
we shall only be concerned with homogeneous, anisotropic and
spatially flat solutions. It is useful to express the metric as
 \ba
ds^2 &=& \exp(\Phi/2)ds^2_E \\
\nonumber
ds^2 &=&\exp(\Phi/2)(-N^2(T)dT^2+\Sigma_{i=1}^{9}
a_E^2(T)e^{2\beta_i}dx_i^2), \ea 
where $ds^2_E$ is the Einstein
frame line element, the Einstein frame scale factor $a_E$ measures the isotropic expansion and the $\beta$'s which are defined such that $\sum_{i=1}^9 \beta_i=0$ measure the anisotropies. The action then takes the form 
\be
S=\frac{1}{2\kappa_{10}^2}\int dT d^{9} x \frac{a_E^9}{N}\left(
-72\frac{{a_E}^2_{,T}}{a_E^2}+\frac{1}{2}{\Phi_{,T}}^2+\sum_i
{(\beta_i)_{,T}^2} \right). \ee Varying with respect to $N$ gives the
constraint \be
-72\frac{{a_E}^2_{,T}}{a_E^2}+\frac{1}{2}{\Phi_{,T}}^2+\sum_i
{{\beta_i}_{,T}}^2=0, \ee and on choosing $N=1$ we obtain the equations
of motion \be \frac{d}{dT}(a_E^9\frac{d\ln
a_E}{dT})=\frac{d}{dT}(a_E^9\frac{d\Phi}{dT})=\frac{d}{dT}(a_E^9\frac{d\beta_i}{dT})=0.
\ee The generic solutions are given by \ba \beta_i(T)
&=&\beta_i(1)+(p_i-\frac{1}{9}-\frac{A}{4})\ln T,
\nonumber \\ \Phi(T) &=& \Phi(1)+A \ln T, \nonumber \\
a_E(T) &=& a_E(1)T^{1/9}, \ea with the following constraints \be
\sum_{i=1}^9 p_i-\frac{9A}{4}=1, \qquad \sum_{i=1}^9 p_i^2-\frac{A}{2}-\frac{A^2}{16}=1. \ee 
We choose to order the $p_i$ such that $p_{i+1} \le p_i$ consistent with our earlier convention. In the
case where $A=0$ we recover the familiar Kasner solutions in the standard
parametrization. If $A>0$ the singularity at $T \rightarrow 0$
occurs at weak coupling as in the ekpyrotic/cyclic models, i.e. $g_s
\rightarrow 0$ \cite{ekpyrotic,cyclic}. If $A<0$ the singularity occurs at strong coupling
as in the pre-big bang model \cite{PBB}. For the pure Kasner solutions
($A=0$) it is easy to see that at least one direction must always be
expanding rather than contracting, implying that there are no fully
isotropic solutions with fixed dilaton. However, allowing the
dilaton to vary, there is then a non-zero set of solutions in which
each direction contracts or remains of fixed size as the
singularity is approached. It is useful to note that all the string
theories have these as solutions to their leading order in $\alpha'$
and $g_s$ equations of motion.

We shall intentionally assume that other contributions such as form fields and curvature terms are sub-dominant contributions to
the space-time dynamics. In practice as we evolve towards the singularity this
remains true for part of the evolution and fails at `Kasner bounces'
where these additional contributions can come to dominate and cause
the system to evolve from one Kasner-like solution to another
\cite{BKL}. We shall also ignore for simplicity their effect on the worldsheet dynamics. These bounces are responsible for the chaotic behavior
associated with spacelike singularities. Our neglect of this may be
justified either by an additional mechanism such as a $w>1$ fluid,
whose attractor behaviour suppresses inhomogenities and can remove
the chaos \cite{Wesley1}, or by topological restrictions that allow us to avoid chaotic behavior until we reach the
string regime where the usual analysis breaks down and we may assume
that quantum string effects remove it \cite{Wesley2}.

Denoting the scale factor that contracts most rapidly by $a_1$ and
that least rapidly or that expands most rapidly by $a_9$, from the previous section we have
two conditions for a well behaved classical evolution towards the
singularity. The first is that the singularity occurs at finite time
in the natural conformal time associated with $a_1$, i.e. where the
metric is \be ds^2=a_1^2(-dt^2+dx_1^2)+ \dots , \ee and the second
is that $a_1^2a_9^2$ should remain finite at the singularity. Let us
consider what this implies for the generic curvature singularity.
The second condition picks out a large class of the Kasner-like
solutions for which
\be
\label{cond1}
p_1+p_9 \ge 0. 
\ee  
In particular if we consider a fully isotropic solution we have $p_i=1/9+A/4$
for $i=1 \dots 9$ and $A \ge -4/9$. Solving the Kasner constraint we
find $A=+4/3$ is the only solution consistent with (\ref{cond1}). This solution is
precisely that considered in ref. \cite{TPS} as a toy model for the
cyclic universe and may be dimensionally oxidized to 2d Milne
$\times R^9$ in M-theory. The second isotropic solution which does not satisfy (\ref{cond1}) describes a pre-big-bang like collapse which occurs at large string coupling. Since loop corrections will inevitably play a significant role in these examples, and the string approach may well be inappropriate we cannot infer anything about these cases with the current approach.

The first condition gives $p_1<1$ and is satisfied for all the Kasner-like solutions except one ($p_1=1$, $p_{i \neq 1}=0$, $A=0$), 
namely when the string frame
metric is 2d Milne $\times R^8$ \be
ds^{2}=-dx_0^{2}+x_0^{2}dx_1^{2}+\sum_{i=2}^{9}dx_i^{2}. \ee
The reason being, in order to put it in the above form we must redefine $%
x^0=-e^{-t}$ with the singularity at $t=\infty $ so that \be
ds^{2}=e^{-2 t}\left( -dt^{2}+dx_1^2\right) +\sum_{i=2}^{9}dx_i^{2}, \ee giving $a_1=e^{-t}$ and $a_9=1$. However, the associated Hamiltonian
describes evolution in $t$ and the singularity is no longer located
at finite $t$ but rather at $t=\infty $. Thus although the
Hamiltonian remains well defined for all finite $t$ all this shows
is that the evolution is well defined before the singularity. 

\subsection{Mode freezing}

\label{PhysImpl}

It is well known that strings on cosmological space-times freeze, i.e. stop oscillating, when the curvature becomes large due to the cosmological expansion or contraction \cite{veneziano2}. In ref. \cite{veneziano2} this was considered by expanding the string solutions around a $\sigma$ independent background. Here we shall consider some exact nonlinear solutions that allow us to see more precisely the condition for the onset of the freezing. Once the modes have frozen, the size of the string scales with the cosmological contraction. 

It is simplest to consider an isotropic contraction/expansion, and use the Polyakov form of the string action
\be
S=\frac{\mu}{2} \int d\tau d\sigma \, \left( (\partial x^0)^2-a(x^0)^2 (\partial \vec{x})^2 \right).
\ee
Here $(\partial A)^2=-(\partial_{\tau} A)^2+(\partial_{\sigma} A)^2$. The equations of motion for $\vec{x}$ are 
\be
\partial(a^2(x^0) \partial \vec{x})=0,
\ee
and the constraints are
\ba
\begin{array}{l}
(\partial_{\tau} x^0)^2+(\partial_{\sigma} x^0)^2-a^2(x^0) \left( (\partial_{\tau} \vec{x})^2+(\partial_{\sigma} \vec{x})^2 \right) = 0 \\
 \partial_{\tau}x^0 \partial_{\sigma}x^0-a^2(x^0) \partial_{\tau}\vec{x} \partial_{\sigma}\vec{x} = 0.
\end{array}
\ea
We look for solutions in which $x^0$ is independent of $\sigma$ and so
\be
\partial_{\tau}^2 \vec{x}+2\frac{a_{,\tau}}{a} \partial_{\tau} \vec{x} -\partial_{\sigma}^2 \vec{x}=0.
\ee
We make the ansatz (see ref. \cite{cosmic} for similar numerical solutions): 
\be
\vec{x}=\vec{x}_0+2\pi \alpha' \int \frac{\vec{p}_0}{a^2} d\tau +\frac{f(\tau)}{a} \left( \vec{c}_1 \sin(n \sigma)+\vec{c}_2 \cos(n\sigma) \right),
\ee
where $f$ satisfies the equation of motion
\be
\label{eqf}
\partial_{\tau}^2 f+n^2 f -\frac{a_{,\tau \tau}}{a} f =0,
\ee
with $\vec{p}_0$, $\vec{c}_1$ and $\vec{c}_2$ are all mutually orthogonal and ${\vec{c}_1}^{\, 2}={\vec{c}_2}^{\, 2}=1$. Here the integer $n$ is the usual mode oscillator number. 
This ansatz is a solution of the equations of motion and the constraints with $x^0$ given by
\be
\frac{dx^0}{d \tau} =2\pi\alpha' \sqrt{\frac{\vec{p}_0^2}{a^2}+M^2(\tau)}
\ee
where we have defined the `mass squared' of the string state as 
\be
M^2(\tau) =\mu^2 \left[  n^2 f^2 +a^2 \left( \frac{f}{a} \right)_{,\tau}^2  \right].
\ee
$\vec{p}_0$ is the comoving momentum so that $\vec{p}_0/a$ is the physical momentum of the particle-like state that the string describes. To begin with, let us assume that $n^2 \gg \left(\frac{a_{,\tau}}{a} \right)^2 , \frac{a_{,\tau\tau}}{a}$ so that $f \sim e^{\pm in \tau}$. Taking the solution
\be
f=\frac{M_0}{\mu \sqrt{2}n}\left( \cos(n\tau) \pm \sin (n\tau)) \right)
\ee
we find that $M^2(\tau) \approx M_0^2$. In this WKB regime we can view the string as describing a particle state with physical momentum $\vec{p}/a$ and mass $M_0$. 
In the quantum theory we expect $M_0^2 \sim 4n/\alpha'$ (for large $n$ where we can neglect the normal ordering constant), although classically it is continuous. As the universe contracts and the curvature increases, at some point the modes will stop oscillating and freeze. The analogue of this is cosmological perturbation theory is when the modes cross the Hubble horizon. 

We have two distinct regimes determined by the magnitude of the particle's momentum:

\subsubsection*{`Short wavelengths'}

As the universe contracts the physical wavelengths blueshift and there is always a regime for which $\lambda_{phys} = 2\pi a/|\vec{p}_0|  \ll l_s/\sqrt{n}$ before the mode freezing sets in (here $l_s=\sqrt{\alpha'}$). In these cases
\be
\frac{dx^0}{d \tau} \approx \frac{\alpha'}{\lambda_{phys}}
\ee
For $n^2 \gg |a_{,\tau\tau}/a|$ the solutions of equation (\ref{eqf}) are oscillatory, however for $n^2 \ll |a_{,\tau\tau}/a|$ they freeze. The criterion for this transition is
\be
n^2 < |a_{,\tau \tau}/a| \approx \left( {\partial_{\tau} x^0 }\right)^2 R_s
\ee
where $R_s \sim 1/l_c^2$ is some measure of the curvature in string frame (not necessarily the Ricci scalar which may vanish). Consequently this is when
\be
n \lambda_{phys} \le l_s^2/l_c.
\ee
Since we are assuming $l_c > l_s$ before the bounce, this will only occur for physical wavelengths less than the string scale which is consistent with the original condition required. 

\subsubsection*{`Long wavelengths'}

If $\lambda_{phys} \gg l_s/\sqrt{n}$ at the onset of the transition, then 
\be
\frac{dx^0}{d \tau} \approx \frac{M_0}{\mu}
\ee
and so the criterion for the transition is
\be
n^2 <  |a_{,\tau\tau}/a| \approx \frac{M_0^2 R_s}{\mu^2}
\ee
which for a quantum string with $M_0^2 \sim n/\alpha'$ gives
\be
n \le \frac{l_s^2}{l_c^2}.
\ee
Thus in this case the effect only occurs when $l_c \approx l_s/\sqrt{n}$, i.e. at the bounce itself. The former momentum dependent behavior is precisely what we obtain in the case of plane waves \cite{Horowitz} and is consistent with the analogy we shall make between the two in the following sections. The latter case is the criterion obtained in \cite{veneziano2}.

We may now see an important distinction between the class I and class II geometries. For class I space-times the singularity is at finite conformal time $t$ and the behavior of the scale factor will typically be $a \sim (-t)^p$ as $t \rightarrow 0$. The ratio of the physical wavelength to the Hubble horizon is
\be
\frac{\lambda_{phys}}{H^{-1}}=\frac{\lambda_{com}}{\mathcal H^{-1}}
\ee
where ${\mathcal H} =p/t$ is the comoving Hubble constant. Since ${\mathcal H}^{-1} \rightarrow 0$ as $t \rightarrow 0$ the majority of modes are stretched beyond the Hubble horizon. Consequently $\lambda_{phys}>l_c>l_s$ before the bounce. Thus in these cases the strings only become excited for $n \sim l_s^2/l_c^2$ i.e. at the bounce itself. For class II geometries the singularity is at infinite conformal time with typically $a \sim t^{-s}$ where $s>1$ as $t \rightarrow +\infty$. In these cases
\be
\frac{\lambda_{phys}}{H^{-1}}=\lambda_{com} \left( -\frac{s}{t} \right) \rightarrow 0,
\ee
at the singularity. Consequently the majority of modes of physical interest will all lie within the Hubble horizon at the bounce. 

One issue we have not addressed in the above analysis is the distinction between the $a_{,\tau ,\tau}/a<0$ behavior and $a_{,\tau \tau}/a>0$. In ref. \cite{veneziano2} the later case was argued to be more unstable (referred to as a Jeans-like instability). This is seen in the fact that the function $f$ grows in magnitude as we approach $a(x^0) \rightarrow 0$. If $a \sim \tau^q$ then the regime $a_{,\tau \tau}/a>0$ corresponds to $q>1$ (and $q<0$ which does not correspond to a singularity at finite worldsheet time). In this cases $f \sim A\tau^q +B \tau^{1-q}$, and it is the second term that diverges. The distinction between class I and class II spacetimes is $q<1/2$ and so the unstable cases all correspond to class II geometries, at least for isotropic solutions. We shall discuss the importance of this for string mode excitation in section \ref{nullcosmological}. Since the solutions discussed above are very special one may be concerned that the transition region inferred from them is not generic. However, a semiclassical analysis predicts the same behavior. Although these solutions may be unstable, the instability does not set until the transition region and consequently does not affect our estimates of when the WKB approximation breaks down. 

\section{From anisotropic cosmologies to plane waves}

\label{spacelike}

In this section we point out an interesting similarity between the behavior of strings on cosmological spacetimes and plane waves. This connection seems to be most clear for anisotropic spacetimes in which one direction contracts more rapidly than the others. In evolving the strings all the way to the singularity in section \ref{classicalstrings1}, we inevitably enter the regime where the strings are effectively tensionless, i.e. all the massive string states are in the ultra-relativistic regime $p_{phys} \gg M_0$, $\lambda_{phys} \ll l_s/\sqrt{n}$. If the geometry bounces at a given scale $l_c$, then in practice it is only the short wavelength modes for which $\lambda_{phys} \ll l_s/\sqrt{n}$ which can be treated as tensionless. The fact that strings become tensionless suggests that the natural expansion parameter should be $\mu$, i.e. $1/\alpha'$ rather that $\alpha'$ near the singularity. This idea has been used in \cite{nullstrings} and more recently in \cite{TPS} in the context of the ekpyrotic/cyclic universe.  If we assume a bounce at the string scale, then because of the arguments of the previous section it is actually only a fraction of the modes for which this is a useful approximation. Furthermore as $n$ becomes large this fraction becomes negligible as one may expect. Nevertheless this is still an interesting regime and in this section we point out that we anticipate the classical and quantum behavior of strings in this regime too be well modeled by strings on plane wave backgrounds.  

\subsection{Semiclassical expansion}

The analogy between strings on cosmological spacetimes and plane waves in the tensionless regime may already be seen at the semiclassical level. A similar semiclassical approach was taken in \cite{semiclassical}. Let us expand the string fluctuations around classical solutions which are
$\sigma$ independent and thus represent particle-like trajectories.
As is usual for the string we have to satisfy the constraint
equation \be
g_{\mu\nu}\partial_{\tau}x_{cl}^{\mu}\partial_{\tau}x_{cl}^{\nu}+g_{\mu\nu}\partial_{\sigma}x_{cl}^{\mu}\partial_{\sigma}x_{cl}^{\nu}
= 0. \ee Together with the equations of motion, these require the
string to move along a null geodesic. Denoting the metric\be \label{metric10}
ds^2=a_1^2(t)(-dt^2+dx_1^2)+\sum_{i=2}^9 a_i^2(t) dx_i^2, \ee and
the string action (in this section $x^0$ denotes conformal time)
\be S=-\frac{\mu}{2} \int d^2\sigma a_1^2(x^0)(-(\partial
x^0)^2+(\partial x^1)^2)+\sum_{i=2}^9 a_i^2(x^0) (\partial x^i)^2,
\ee let us consider geodesics in the $x_1$ direction. We have \be
(\partial_{\tau}x^0)^2-(\partial_{\tau}x^1)^2=0, \ee and so
$\partial_{\tau}x^1=\eta\partial_{\tau}x^0$ with $\eta=\pm 1$
depending on the direction of the trajectory. Since
$\partial_{\tau}(a_1^2(x^0)\partial_{\tau}x^0)=0$ our chosen classical
solution is \ba \nonumber
&& \int a_1^2(x^0)dx^0 =A\tau+B \\
\nonumber
&& x^1 = \eta x^0+C \\
&& x^i = 0 \qquad i=2 \dots 9.
 \ea
 It is convenient to choose coordinates so that $C=0$. Then
 expanding $x^{\mu}=x^{\mu}_{cl}(\tau)+Y^{\mu}$ to second order we
 find
 \ba
 &&S = -\frac{\mu}{2} \int d^2 \sigma a_1^2(x^0_{cl})(-(\partial
Y^0)^2+(\partial Y^1)^2) \\ 
\nonumber
&&+2(a^2)_{,{x^0}}Y^0(\partial
x^1_{cl}\partial Y^0-\partial x_{cl}^0 \partial Y^1)+\sum_{i=2}^9 a_i^2(x^0_{cl}) (\partial Y^i)^2. \ea Let us now
define $u=\frac{1}{2}(x^0+\eta x^1)$ and $v=2(\eta x^1-x^0)$ and
similarly split $u=u_{cl}+\hat{U}$ and $v=v_{cl}+\hat{V}$ with the
classical value given by $u_{cl}=x_{cl}^0$ and $v_{cl}=0$. 
Then the above
action may be expressed as 
\ba 
S&=& -\frac{\mu}{2} \int d^2 \sigma
2a_1^2(u_{cl})(\partial \hat{U}\partial \hat{V}) \nonumber \\
&&+2a^2_{,u}\partial
u_{cl}\partial \hat{U}+\sum_{i=2}^9 a_i^2(u_{cl}) (\partial
Y^i)^2.
\ea 
This is precisely the same action we obtain if we start with the following action \be
S=-\frac{\mu}{2}\int d^2 \sigma  2a_1^2(u) \partial u
\partial v + \sum_{i=2}^9 a_i^2(u) (\partial x^i)^2, \ee 
and perform the same quadratic expansion $u=u_{cl}+\hat{U}$ and $v=v_{cl}+\hat{V}$. Since
we are expanding around a null geodesic, at leading order in the
background field expansion, strings on the metric (\ref{metric10}) behave
identically to strings on the metric \be
ds^2=2a_1^2(u)dudv+\sum_{i=2}^9 a_i^2(u) dx_i^2. \ee 
This metric is a plane wave metric. This is closely
connected to the fact that the second metric is the Penrose limit of
the first. Writing $t=u$, $x_1=u+2\epsilon v$ and $x_i=\epsilon
\hat{x}_i$, the first metric is 
\ba \nonumber
&&ds^2=\epsilon^2a_1^2(u)(-du^2+d(u+2\epsilon
dv)^2)+\sum_{i=2}^9 a_i^2(u) \epsilon^2 d\hat{x}_i^2 \\
&=&\epsilon^2 \left[2a_1^2(u)dudv+\sum_{i=2}^9 a_i^2(u) dx_i^2
\right]+4\epsilon^4 a_1^2(u)dv^2 ,
\ea rescaling $ds^2 \rightarrow
ds^2/\epsilon^2$ and taking the limit $\epsilon \rightarrow 0$ we
discover the plane wave metric. In the first case we are expanding
around a null geodesic and in the second we are boosting infinitely
along that same null geodesic. Although slightly different
procedures they have essentially the same outcome.

\subsection{Near the singularity}

\label{neartosingularity}

We have already seen in section \ref{classicalstrings1} that as we
approach the singularity, classical strings freeze and begin to move
along null geodesics. This fact allows us to make a close connection
between cosmological and plane wave space-times near the
singularity. Consider again the Hamiltonian for the string in proper
time gauge 
\begin{eqnarray} 
\begin{array}{l}
 H= \int d\sigma \\ 
  \sqrt{{P_1}^{2}+\mu^{2}a_1^{4}(t)
\left| x_{1}^{\prime }\right| ^{2}+\sum_{i=2}^9 \left( \frac{1}{r_{i}^{2}}P _{i}^{2}+a_1^2a_{i}^{2} \left| x_{i}^{\prime}\right|^{2} \right)}. 
\end{array} 
\end{eqnarray}
Now by definition as we approach the singularity the Hamiltonian
asymptotes to $H \rightarrow \int d\sigma |P_1|$ and the
leading order nontrivial dynamics in the '$x_{i}$' directions can be
found by Taylor expanding the square root in powers of $1/|P_1|$. Let us
denote $H=H_0+H_1+\dots$ where \be H_0=\int d\sigma \left[
|P_1|+\frac{1}{2|P_1|}\mu^{2}a_1^{4}(t)
\left| x_1^{\prime }\right| ^{2} \right] \ee and \be H_1=\int
d\sigma\frac{1}{2|P_1|}\sum_{i=2}^9\left(\frac{1}{r_{i}^{2}}P_{i
}^{2}+ \mu^{2}a_1^{2}(t) a_{i}^{2}\left|
x_{i}^{\prime }\right|^{2}  \right). \ee What is
remarkable is that $H_1$, which governs the evolution in the
transverse $x_{i}$ directions is precisely the Hamiltonian for strings in
the plane wave space-time. Here we identify $|P_1|=p_{V}$ (defined in appendix \ref{null1}) where $p_V$ is the conserved momentum associated with the null Killing vector which shall play a prominent role in what follows.
This is a straightforward consequence of the evolution of the string
tending asymptotically to a null geodesic. Then the action near
the singularity is 
\ba
\begin{array}{l}
S = \int dtd\sigma \left[ \sum_{i=1}^9 P_i \frac{dx_i}{dt}-(H_0+H_1+\dots) \right] \\
= \int dtd\sigma   P_1
\frac{dx_1}{dt}-\left(|P_1|+\frac{1}{2|P_1|} \mu^{2}a_1^{4}(t) \left|
x_{i}^{\prime }\right| ^{2} \right)   \\
 +\sum_{i=2}^9  P_i \frac{dx_i}{dt}
-\frac{1}{2|P_1|} \sum_{i=2}^9 \left( \frac{1}{r_{i}^{2}}P_{i
}^{2}+ \mu^{2} a_1^{2}(t) a_{i}^{2}\left|
x_{i}^{\prime}\right|^{2}   \right)  .
\end{array}
\ea 
On
defining $V=x_1-\eta t$ we get (here $\eta=P_1/|P_1|$ and has the same meaning as in the previous section)
\ba S&=&\int dtd\sigma P_1
\frac{dV}{dt}-\left(\frac{1}{2|P_1|}\mu^{2}a_1^{4}(t) \left|
V^{\prime }\right| ^{2}
\right) \\
\nonumber
&&+\sum_{i=2}^9  P_i \frac{dx_i}{dt}-\frac{1}{2|P_1|} \sum_{i=2}^9 \left( \frac{1}{r_{i}^{2}}P_{i
}^{2}+ \mu^{2}a_1^{2}(t) a_{i}^{2}\left|
x_{i}^{\prime }\right|^{2}  \right) ,
\ea 
which
makes it clear that the momentum conjugate to $V$ is $p_V=P_1$. If
we consider modes which do not oscillate in the $x_1$ direction we
may neglect the second term in the action and so we have \ba S&=&\int
dtd\sigma P_1
\frac{dV}{dt}+\sum_{i=2}^9  P_i \frac{dx_i}{dt}\\
\nonumber
&&-\frac{1}{2|P_1|}\sum_{i} \left[ \frac{1}{r_{i}^{2}}P_{i
}^{2}+ \mu^{2}a_1^{2}(t) a_{i}^{2}\left|
x_{i}^{\prime }\right|^{2}  \right] +\dots, \ea 
which is
precisely the canonical action of a string on a plane wave
space-time. So it is the modes that do not oscillate in the $x_1$
directions for which the correspondence holds most closely. There is
one important difference, namely that in the null case $p_V$ is
$\sigma$ independent and is a conserved charge of the string
associated with the null Killing vector, whereas in the spacelike
case $|P_1|$ is in general $\sigma$ dependent. Nevertheless
the dynamics is sufficiently similar that we expect a close
correspondence in the solutions.

\section{String mode excitation}

\label{nullcosmological}

In the previous section we have seen that there is a close analogy between strings on anisotropic cosmological spacetimes and plane waves. Since the latter are exact conformal field theories we may use this analogy to anticipate the properties of the quantum string. In ref. \cite{Horowitz} it was observed that on plane wave space-times there is a new physical effect not present for strings on Minkowski space-time; strings may become excited as they propagate in the time-dependent spacetimes. The physical consequence of this is particle transmutation \cite{semiclassical}, a particle of given mass and spin may emerge as a particle with different mass and spin. This effect was further studied in \cite{Sanchez2} and from a semiclassical perspective in \cite{semiclassical}. In appendix A we give a brief review of some of the properties of plane waves and the description of string mode excitation. In appendix B we show that for plane wave solutions, which are conformal field theories, it is impossible for the Einstein frame to bounce. We cannot use this to infer anything about cosmological space-times however since the analogy between them works only for modes for which $\lambda_{phys} <l_s/\sqrt{n}$. As $n \rightarrow \infty$ this becomes increasingly difficult to satisfy. The renormalization group equations depend on the UV properties of the string and hence on the $n \rightarrow \infty$ behavior. 

The metric of the plane wave in cosmological form is
\be \label{metric1} ds^2=2a_1^2(u) dudv+\sum_{i=2}^9 a_i^2(u)dx_i^2,
\ee
and may be converted to the more familiar Brinkmann form by defining 
$U=\int a_1^2(u) du$, $X_i=a_i(u)x_i$ and similarly
$V=v-\frac{1}{2}\sum_i \frac{{a_i}_{,U}}{a_i}X_i^2$ so that
\be ds^2=2dUdV+\sum_i \frac{{a_i}_{,UU}}{a_i}X_i^2dU^2+dX_i^2. \ee
Using the notation of appendix A, the string mode functions satisfy
\be
\ddot{\psi}^i_n+n^2\psi^i_n-\left(\frac{p_V}{\mu} \right)^2k_{ii}\left(\frac{p_V \tau}{\mu} \right)\psi^i_m=0.
\ee 
where $k_{ii}=a_{i,UU}/a_i$. This may be thought of as a non-relativistic scattering problem, where the last term is the scattering potential and $n^2$ plays the role of the energy. In ref. \cite{Horowitz} a generic argument for the behavior of the modes was given based on the dimensionless ratio $n\epsilon_i \mu/p_V$ where $\frac{1}{\epsilon_i}$ is the maximum value of $\sqrt{|k_{ii}|}$. If $n \gg p_V/\mu\epsilon_i$ then the scattering potential is small and the Born approximation may be used. The Bogoliubov coefficients describing the amount of mode excitation are given by 
\be
\beta^i_n=-i\tilde{k}_{ii}(\frac{n}{\pi\alpha'p_{V}})\frac{\pi\alpha'p_V}{n},
\ee
where $\tilde{k}_{ii}$ is the Fourier transform of $k_{ii}$ and they have the form
\be
\beta^i_n \sim \frac{\pi^2\alpha'p_V}{n\epsilon_i}\exp\left(-\frac{n\epsilon_i}{\pi\alpha'p_V} \right).
\ee
Associated with each direction we may define a curvature scale or Hubble horizon as
\be
\frac{1}{l_c^i} =\left(  \frac{a_{i,u}}{a_1 a_i} \right)=a_1 \frac{a_{i,U}}{a_i}.
\ee
Translating into our previous notation we find that in the regime $n \gg p_V/\mu\epsilon_i$ is equivalent to
\be
n \lambda_{phys} \ll l_s^2/{\text{Min}}(l_c^i),
\ee
where $\text{Min}(l_c^i)$ is the minimum curvature length and $\lambda_{phys}$ is the wavelength at the time the maximum curvature is reached, i.e. at the bounce. If the bounce occurs at the string scale then this requires $\lambda_{phys} \ll l_s/n$. However the plane wave analogy is only valid if $\lambda_{phys} \gg l_s/\sqrt{n}$ and so this result is not appropriate. We may improve on this analysis by considering semiclassical perturbations around classical configurations describing massive states
\cite{semiclassical}, however it is straightforward to see intuitively how the result will behave from the analysis of section \ref{PhysImpl}. When $\lambda_{phys} \gg l_c/\sqrt{n}$ up until the bounce, then the criterion for the modes to freeze is $l_c^i \approx l_s/\sqrt{n}$. The Bogoliubov transformation is associated with the onset of the freezing, i.e. when the WKB approximation breaks down. Thus we anticipate that in this regime the $\beta$'s to behave as
\be
\label{beta1}
\beta_n^i \sim a \frac{l_s}{\sqrt{n}l_c^i} \exp \left(-b \sqrt{n} \frac{l_c^i}{l_s} \right),
\ee
where $a$ and $b$ are dimensionless coefficients of $O(1)$. Strictly speaking this is too simplistic since unlike for plane waves, in the cosmological case the different coordinates of the string interact and it is no longer appropriate to think of independent Bogoliubov coefficients for each direction $X^i$. This interaction will spread out the amount of excitation between the different excited states. However, in terms of estimating the overall amount of excitation, it is likely that (\ref{beta1}) gives a reasonable estimate of the probability to excite a mode $n$. This falls of only as an exponential of $\sqrt{n}$ is because we have assumed the mass of the string increases as $\sqrt{n}$. This result shows that the amount of mode excitation on a cosmological space-time will be in general be greater than that anticipated for their plane wave counterparts, where the fall off is $\exp(-n\lambda_{phys} l_c^i/l_s^2))$. In the latter case even for $n \sim 1$ and $l^i_c \sim l_s$ for modes whose wavelengths are much greater than the string scale, this amplitude is exponentially small. However, in the cosmological case for $\lambda_{phys}>l_s/\sqrt{n}$ we expect no suppression for large wavelength modes, and we expect an $O(1)$ probability of excited states at all wavelengths. There may however be an additional suppression, for instance if the coefficient $b$ turns out to be reasonably large. This depends on precisely what happens at the bounce, thus it is conceivable that depending on the string scale physics this effect could be significantly suppressed or enhanced.

At small wavelengths $\lambda_{phys} \ll l_s/\sqrt{n}$ we may trust the analogy with plane waves. In ref. \cite{Horowitz} a WKB approximation was use to study the regime $n \ll p_V/\mu\epsilon_i$ i.e. $n\lambda_{phys} \ll l_s^2/l_c^i$ assuming that the time-dependent plane wave regime was bounded by two regions of Minkowski spacetime. In the present context this approximation is not valid. However, we may now use the intuition developed in section \ref{classicalstrings1}, to see what happens in each class. A given mode will freeze when $\lambda_{phys} \approx l_s^2/nl_c^i$ which occurs at a curvature scale $l_c^i \gg l_s /\sqrt{n}$. Conseqently this will occur well before the bounce, in the regime in which the collapse is governed by the low energy supergravity equations. Once the modes have frozen they track the contraction and subsequent expansion of the scale factor
\be
\psi^i_n \approx C_1 a_i+C_2 a_i \int \frac{1}{a_i^2} d\tau.
\ee
For class I spacetimes in which the integral is finite as $a_i \rightarrow 0$ the subsequent evolution is largely independent of how $a_i$ is regulated near the bounce. For class II spacetimes the integral is divergent as $a_i \rightarrow 0$ and so will be strongly sensitive to the manner in which it is regulated, i.e. precisely what happens at the bounce. In practice this means that the Bogoliubov coefficients in the regime $\lambda_{phys} \ll l_s/\sqrt{n}$ are largely insensitive to $\epsilon_i$/$l_c^i$ whereas for class II spacetimes they are strongly sensitive. 

We may construct a toy analytic model of a bouncing space-time as follows: Consider a solution in which for the different regions of $U$, $a(U)$ takes the form
\ba
a(U)=(-U)^{\nu} \qquad &&U \le-\epsilon ,\\
a(U)= \epsilon^{\nu} \frac{\cosh\left( \frac{qU}{\epsilon} \right)}{\cosh q}  \qquad &&-\epsilon < U< \epsilon ,\\
a(U)= U^{\nu} \qquad &&U \ge \epsilon,
\ea
where $q \tanh q=\nu$. 
This describes a scale factor which contracts as a power law, bounces and then expands as a power law. We can distinguish the different class by the power $\nu$. In this example $\nu <1/2$ corresponds to class I space-times and $\nu>1/2$ to class II space-times. The factor of $\cosh(q)$ in the intermediate solution arises by ensuring that $a(U)$ and $a'(U)$ are continuous at $U=\pm \epsilon$. 

A solution which describes an incoming positive frequency mode as $u \rightarrow -\infty$ takes the form
\ba
\begin{array}{l}
\psi^+(\tau) = \sqrt{-n\tau} H^{(1)}_{\nu-1/2}\left(n \tau \right), \qquad U \le -\epsilon \\
= C_1 e^{ \tau \sqrt{\left( \frac{p_Vq}{\epsilon \mu}\right)^2-n^2}}+C_2 e^{ -\tau \sqrt{\left( \frac{p_Vq}{\epsilon \mu}\right)^2-n^2}},  \, \, -\epsilon < U < \epsilon \\
= \alpha  \sqrt{-n\tau} H^{(1)}_{\nu-1/2}\left(n \tau \right)
+\beta \sqrt{-n\tau} H^{(1)}_{\nu-1/2}\left(n\tau \right), \, \, \, U \ge \epsilon.
\end{array}
 \ea
The constants $C_1, C_2, \alpha, \beta$ may be determined by matching the field and its derivative at $u=\pm \epsilon$. On computing the full expression for the Bogoliubov coefficients we find that for $\nu<1/2$ in the limit $n\epsilon/p_V\mu \rightarrow 0$
\be
\beta^i_n \rightarrow -i \cot\left((\nu-1/2) \pi \right),
\ee
which is precisely the result obtained by matching the field and its momentum conjugate at the singularity according to the prescription of section \ref{classicalstrings1}. Note that this is independent of $n$ and $\lambda_{phys}$ and is inevitably of $O(1)$. The same argument applied to the case $\nu > 1/2$ gives the diverging behavior
\be
\beta^i_n \sim \left( \frac{\alpha' p_V}{n\epsilon}\right)^{1-2\nu}.
\ee
This becomes arbitarily large at wavelengths $\lambda_{phys} \ll l_s/n$ assuming a string scale bounce.  Note that the two qualitative behaviors are not distinguished according to whether $a_{,UU}/a >0$ or $a_{,UU}/a<0$ as considered in \cite{Horowitz,veneziano2} since these regimes correspond to $\nu>1 $ or $\nu<0$ and $0<\nu<1$. It appears then than the Jeans-like instability is less important that whether the space-time belongs to class I or class II. In this sense then we find that even at the quantum level the two classes of space-time are distinguished by the sensitivity of the string propagation to the bounce.

\section{Implications for Ekpyrotic/Cyclic models}

\label{ekpyrotic/cyclic}

Let us now consider the implication of these results for realistic cosmologies that incorporate the idea of a bounce such as the ekpyrotic and cyclic universe models. The issue we need to address is what does the   excitation of strings imply for the propagation of cosmological perturbation through a string scale bounce. Cosmological perturbations on bouncing universes have been considered in several different ways in the literature \cite{perts}. Here we shall be concerned with what happens as the perturbations evolve through the bounce. A gravitational wave (tensor perturbation) is a coherent state of gravitons. The results of the previous section imply that if certain conditions are met, there is significant probability that an incoming graviton emerges from the bounce as an excited string state. In field theory terms, the incoming gravitational wave perturbations become converted into the perturbations of several string scale mass fields. A similar story holds for scalars and one thus ends up with non-adiabatic perturbations. However, in a short time scale these will decay back to low mass/massless perturbations. It may be possible to model this process by an effective action including the excited string states and various interactions between them and the massless states. This would allow us to get a better handle on how the cosmological power spectrum is modified by this intermediate string regime. However, recent work \cite{Wesley3} suggests that a field theory approach may not be an adequate description. If this is the case, then only a fully string calculation of the propagation of cosmological perturbations will resolve this issue. At the present time this seems to be beyond the current technology. 

The ekpyrotic/cyclic models are based on the collision of orbifold planes (end of the world branes) in heterotic M-theory. Since $g_s \rightarrow 0$ as the orbifold planes approach, we expect 10d heterotic string methods to be appropriate. In the simplest models of these collisions, the CY moduli are assumed to be stabilized from the 11d perspective. Consequently in 10 dimensions the string frame metric will contract isotropically
\be
ds^2=a_s^2(-dt^2+d\vec{x}^2+ds^2_{CY}),
\ee
where $ds^2_{CY}$ is the Calabi-Yau metric. As the geometry collapses, modes are pushed outside the horizon ($\lambda_{phys}/H^{-1} \rightarrow \infty$). When the string frame curvature becomes of order $R_s \sim 1/\alpha'$ we can no longer trust the low energy supergravity. Nevertheless if modes have become excited already before this scale then these will have some impact on the late time phase. 
In this model long wavelength perturbations are generated by modes which have crossed the horizon in the collapsing phase. Thus these will have $\lambda_{phys} \gg l_s$ at the bounce. For these wavelengths we may expect to excite with a probability of $O(1)$ the first few excited string states $n=1,2 \dots$ at all wavelengths $\lambda_{phys} \gg l_s$. The arguments of the previous section seem to indicate that this probability is independent of wavelength. This is interesting since it suggests that the momentum dependence of the incoming perturbation spectrum  will be unchanged, although the amplitude will be. However, this depends on precisely how the excited states decay back to the massless/low mass states. The momentum independence is likely to be a result of the fact that long wavelength modes have already frozen in (in the usual space-time sense) before we reach the bounce and so all spatial derivatives of the fields may be neglected. This is connected with the usual ultralocal behavior near cosmological singularities \cite{BKL}.

The caveat in these conclusions is that as we have seen, for modes with $\lambda_{phys} \gg l_s$ they become excited when the curvature scale is close to the string scale and it is precisely in this regime that string $\alpha'$ corrections kick in and we can no longer trust the preceding analysis. Whether the string excitation remains significant or is somehow suppressed would require a deeper understanding of the string scale physics, however we feel that qualitatively the resulting physics will be similar. As well as particle transmutation, we also expect particle/string creation on cosmological space-times. Generic estimates of string production have been given in \cite{Gubser} in an effective field theory approach and more specifically for the ekpyrotic/cyclic in ref. \cite{TPS} via an instanton approach. The validity of both these calculations in the string regime is unclear \cite{Wesley3}, and more work needs to be done on understanding these effects and the implications of the backreaction of created particles. 

In the regime $\lambda_{phys} \ll l_s$ things are under better control since the strings become excited before we reach the string scale. Our analysis suggests a significant amount of excitation of the string at very short wavelengths. Since this effect occurs well before the string scale we expect this result to be largely unmodified by the physics of the bounce. Cosmologically these modes are far less of a problem. Again we anticipate they will decay in the expanding phase and may give rise to small scale density perturbations or the formation of small black holes which would ultimately radiate away.

In ref. \cite{TPS} a proposal was made to match strings and membranes across the collision of two orbifold planes in heterotic M theory. From the 10d perspective these winding modes correspond to strings, and the fact that the strings may be continued across the singularity is simply a consequence of the fact that $a_s^2 \sim (-t)$ as $t \rightarrow 0^-$ and so this geometry corresponds to a class I spacetime for which we may directly match the comoving position and momentum of the string across the singularity. If we evolve the strings all the way up to $t=0$ as advocated in ref. \cite{TPS}, then we find that all the modes of the string freeze and begin to move at the speed of light at the collision. If we consider the Penrose limit of this geometry, so that $a^2_s \sim (-u)$, we may exactly solve the classical and quantum string mode functions all the way up to the singularity, and directly match them across into the expanding phase. As in the cosmological case, the classical solutions are well-behaved and no problem arises in the prescription. However, in the quantum case, although there is no technical problem with matching the position and momentum operators, on calculating the Bogoliubov coefficients we find that the ${\beta_n}'s$ are constants and independent of $n$, i.e. each mode on the string is excited with equal probability. Since there are an infinite number of modes on the string, the total excitation of the string and its mass are infinite and so the S-matrix theory breaks down. What this shows is that even when the classical behavior of strings is well behaved, the quantum behavior may be pathological. Whether this remains true in the cosmological case is not clear without a rigorous quantization, but in the previous sections we have argued that very similar physics takes place.

\section{Summary}

We have seen that generic (an)isotropic spatially flat cosmological singularities may be divided into two classes from the perspective of the propagation of classical strings. In class I, the strings are well behaved all the way to the singularity. In class II they become ill-defined at the singularity. We have seen that part of this feature remains in the quantum theory, at least with regards to the excitation of string modes, but that the situation is more subtle and depends on the momenta of the string states. 
Quite independently of whether the singularity is resolved by a bounce or some other string scale state, the distinction between the worldsheet string physics of the two classes deserves further exploration. We have reconsidered the issue of string mode excitation (particle transmutation) through a bouncing cosmology. We have argued that the amount of mode excitation depends qualitatively on whether the string describes a short or long wavelength mode, in comparison to the string scale at the bounce. In the former case the physics is well under control and a close analogy may be made to strings on plane wave space-times. For long wavelength modes things are less well under control since the excitation occurs in the stringy regime near the bounce. Nevertheless taking the semi-classical arguments seriously suggests that modes of all wavelengths will emerge in mildly excited states which would have important cosmological implications. A better understanding of the string scale regime must be developed in order to fully understand the cosmological implications of the bounce.

\section*{Acknowledgements}

I would like to thank Daniel Baumann, Justin Khoury, Gustavo Niz, Claudia de Rham, Neil Turok, Dan Wesley, Toby Wiseman and in particular Paul Steinhardt for useful comments. 
This work was supported
in part by US Department of Energy grant DE-FG02-91ER40671.

\newpage

\setcounter{equation}{0}
\renewcommand{\theequation}{\thesection. \arabic{equation}}

\appendix

\section{Properties of Plane waves}

\label{null1}

Plane wave space-times are special cases of pp-wave space-times
which have the general form \be ds^2=2dUdV+K(U,X)dU^2+dX_i^2 \ee
where $K$ is an arbitrary function of $U$. Plane wave space-times are special symmetric solutions for
which $K(U,X)=k_{ij}(U) X^iX^j$. 
We can generalize the pp-wave solutions to include a non-zero
Kalb-Ramond field. If we assume $B^{(2)}$ has the form \be
B_{iu}=B_i(U,X) \ee with all other components vanishing, then the
leading order in $\alpha'$ renormalization group equations are
reduce to 
\ba 
\label{RG2}-\frac{1}{2}\nabla^2 K-\frac{1}{2}H_{(2)}^2 +2\partial_u^2 \Phi &=&0 \\
\partial^iH_{ij}&=&0,
\ea with $H_{ij}=\partial_i B_j-\partial_j B_i$ and
$H_{(2)}^2=\frac{1}{2!}H_{ij}H^{ij}$. If we make the further ansatz
$B_i=-\frac{1}{2}H_{ij}(U)X^j$, then the string sigma model is 
conformally invariant to all orders in $\alpha'$. This is simply
because for this form, $B^{(2)}$ contributes a quadratic term in
$X^i$ to the string action. Again specifying $K=k_{ij}(U)X^iX^j$ these define the so-called Hpp-waves.

One of the interesting aspects about these null space-times is the
apparent reduction of constraints in comparison to the spacelike
cases. For instance, there is no nontrivial equation of motion for
$\Phi(u)$ since \be \Box \Phi=(\partial \Phi)^2=0 \ee are
automatically satisfied since $g_{UU}=0$. This gives us freedom to choose
$k_{ij}(U)$, and we may still satisfy the renormalization group
equations by taking the dilaton to satisfy (\ref{RG2}). 
This should be contrasted with spacelike cosmologies,
where we have an additional equation for $\phi(t)$. This extra
freedom can be understood in terms of the differing nature of the
Cauchy problem. In the spacelike case, a surface $t=constant$ is a
Cauchy surface, whereas in the null case $u=constant$ is not a
complete Cauchy surface and we must impose additional data on say a
$V=constant$ hypersurface to completely specify the solution. It is this additional freedom that allows
us free choice for $k_{ij}(U)$ or equivalently $\Phi(U)$.

These geometries are known to be exact conformal field theories
provided the equation \ref{RG2} is satisfied. In the case of plane
waves this may be shown by an exact calculation of the path integral
\cite{Amati}, however it may equivalently be seen to arise from the
fact that all higher order contractions of the Riemann tensor and
with the dilaton and Kalb-Ramond 3-form flux vanish, and so the
usual $\alpha'$ corrections to the leading supergravity equations of
motion vanish \cite{Horowitz}. 

\subsection{Calculation of S-matrix}

The action for the string on a plane wave metric in Brinkmann form is
\ba 
\begin{array}{l}
S=-\frac{\mu}{2}\int d^2 \sigma 
 \eta^{ab} (
2\partial_aU\partial_bV   \\  +\partial_aX^i\partial_b X^i+k_{ij}(U)X^iX^j
\partial_a U\partial_b U     ) .
\end{array}
\ea From the equation of motion $\partial^a\partial_aU=0$ we may set
$U=2\pi\alpha'p_V\tau$ giving the action 
\ba 
\begin{array}{l}
S= \int d^2 \sigma
(-p_V\partial_{\tau}V  \\ -\frac{\mu}{2} (\eta^{ab}
\partial_aX^i\partial_b X^i- k_{ij}(U)X^iX^j(\frac{p_V}{\mu})^2
)) ,
\end{array}
\ea 
and so the equation of motion for the $X^i$ are \be
\partial_{\tau}^2X^i-\partial_{\sigma}^2X^i-\left(\frac{p_V}{\mu} \right)^2
k_{ii}\left(\frac{p_V}{\mu} \tau \right)X^i=0, \ee 
where we assume $k_{ij}$ is
diagonal or has been diagonalized. The general solutions to this
equation may be expressed as 
\ba X^i &&=
X_0^ia^i+(2\pi\alpha')P_0^ia^i\int\frac{1}{a_i^2}dU \\ \nonumber &&
-i(\frac{\alpha'}{2})^{1/2}\sum_{m=-\infty,\neq0}^{\infty}\left(
\frac{\alpha_m^i}{m} \psi^i_m e^{-im\sigma} +
\frac{\tilde{\alpha}_m^i}{m} \psi^i_m e^{im\sigma} \right), \ea such
that $\psi^i_{-m}=(\psi^i_m)^*$ and \be
\ddot{\psi^i_m}+m^2\psi^i_m-\left(\frac{p_V}{\mu} \right)^2k_{ii}\left(\frac{p_V \tau}{\mu} \right)\psi^i_m=0.
\ee 
To quantize the system we replace
$(X_0^i,P_0^i,\alpha_m^i,\tilde{\alpha}_m^i,p_V)$ by operators and the
canonical commutation relations imply \ba \nonumber [X^i_0,P^j_0] &=
& i\delta_{ij} \\ \left[ {\alpha_m}^i,{\alpha_n}^j
  \right] &=& \left[ {\tilde{\alpha}_m}^i,{\tilde{\alpha}_n}^j
  \right]  = \delta_{m+n}\delta_{ij},
\ea where we have chosen the normalization so that \be
i({\psi^{i \, *}_m}{\dot{\psi}}^{i}_m-{{\dot{\psi}}^{i \, *}_m} \psi^i_m)=2m.
\ee We now face an ambiguity in the choice of positive frequency
modes $\psi^i_m$ for $m>0$ which exactly parallels the ambiguity in
the choice of vacuum that arise in quantum field theory on a curved
space-time. As $U \rightarrow \pm \infty$ we expect $\psi^i_m
\approx Ae^{-im\tau}+Be^{im{\tau}}$ and so  we naturally define an
$\it{'in'}$ and $\it{'out'}$ vacuum by the requirement that the
modes are positive frequency in the past and future \ba \lim_{\tau
\rightarrow -\infty} \psi^{in,i}_m(\tau) &\approx&
e^{-im\tau}, \\
\lim_{\tau \rightarrow +\infty} \psi^{out,i}_m(\tau) &\approx&
e^{-im\tau}. \ea As in quantum field theory on a curved space-time,
there will be a Bogoliubov transformation between the {\it in} and {\it out}
states \be \psi^{in,i}_m=\alpha_m \psi^{out,i}_m+\beta_m
{\psi^{out,i}_m}^* .\ee Given a state $|i,in \rangle$ which lies on the Fock
space built on the vacuum $|in \rangle$ and a state $|j,out \rangle$ similarly
defined we may construct the S-matrix \be S_{ji}=\langle j,out|i,in \rangle. \ee

\subsection{Physical Implication of Bogoliubov Transformations}

In quantum field theory on curved spaces, the Bogoliubov
coefficients measure the number of particles seen by an {\it out}
observer in the {\it in} vacuum. Specifically if
$N_{out}(k)=b_k^{\dagger}b_k$ is the number operator for an {\it
out} observers definition of particles, then \be
\langle in|N_{out}(k)|in \rangle=|\beta(k)|^2. \ee The total number of particles
seen by an out observer is \be N^{total}_{out}=\int
\frac{d^3 k}{(2\pi)^3}|\beta(k)|^2 .\ee In the present context the
interpretation is similar. The $|in \rangle$ state may be viewed as a
squeezed state \cite{squeeze1,squeeze2} \be |in \rangle=\prod_n
(1-|\gamma_n|^2)^{1/4}\exp\left(-\frac{1}{2} \gamma_n
b_n^{\dagger}b_{-n} \right) |out \rangle,\ee where
$\gamma_n=-\beta_n/\alpha_n^*$. So the $|in \rangle$ state is a
superposition of excited $|out \rangle$ states \be \sum
b^{\dagger}_{n_1}b^{\dagger}_{-n_1}b^{\dagger}_{n_2}b^{\dagger}_{-n_2}
\dots |out \rangle.\ee Note that we always apply the creation operators in
pairs. Now in the context of strings \ba \nonumber
b^{\dagger}_n&=&\frac{\alpha^{out}_{-n}}{n} \\
b^{\dagger}_{-n}&=&\frac{\tilde{\alpha}^{out}_{-n}}{n} \quad {\rm
for} \, n>0.\ea 
Thus 
\be
|in \rangle=\prod_{i,n_i} 
(1-|\gamma^i_{n_i}|^2)^{1/4}\exp\left(-\frac{1}{2n_i^2}
\gamma^i_{n_i}
\alpha^{i,out}_{-{n_i}}\tilde{\alpha}^{i,out}_{-{n_i}} \right)
|out \rangle,
\ee and so we always excite an equal number of left and right
movers relatively to the $|out \rangle$ state, this is necessary in order
to conserve worldsheet momentum. If the plane wave geometry is
asymptotically Minkowski as $U \rightarrow \pm \infty$, then we may
use the standard definitions for physical states in the asymptotic
regimes. For an ${\it{out}}$ observer the mass of the string states
is measured by 
\be M^2_{out}=4 \sum_{i=2}^9 \sum_{n_i=-\infty, \neq 0}^{\infty} n_i b^{\dagger
i}_{n_i}b_{n_i}^i -8, 
\ee 
and so 
\be \langle in|M^2_{out}|in \rangle=
4 \sum_{i=2}^9 \sum_{n_i=-\infty, \neq 0}^{\infty}n_i |\beta^i(n_i)|^2 -8.
\ee Consequently if the
$\beta$'s are large, the ${\it{out}}$ observer measures a large mass
for the string in accordance with its interpretation of the string
as a superposition of massive excited states. If $\langle in|M^2_{out}|in \rangle$
is significantly large, this suggests that the excited strings could
significantly backreact on the geometry. This has been used as a
criterion to asses whether string propogation across null
singularities makes sense at the quantum level \cite{Horowitz,Sanchez2}. 
\setcounter{equation}{0}

\section{Einstein frame and a no-bounce theorem}

\label{Einstein}

For plane waves of the form $k_{ij}(U)=a_i''/a_i \delta_{ij}$ (in this section $'=\partial_U$) the renormalization group equations for the string are
\be -\sum_i
\frac{a_i''}{a_i}-\frac{1}{2}H_{(2)}^2+2 \partial_u^2 \Phi =0.\ee
Let us denote $a_i=a_se^{\beta_i}=a_Ee^{\beta_i}e^{\Phi/4}$ such
that $\sum_i \beta_i=0$. Here $a_s$ and $a_E$ are the string and
Einstein frame scale factors defined by the 8'th root of the spatial
volume element in each frame. After some rearrangements we find the
equations in Einstein frame to be \be
H_E'=-\frac{1}{16}(\Phi'+4H_E)^2-\frac{1}{8}\sum_i{\beta_i'}^2-\frac{1}{16}H_{(2)}^2
\ee where $H_E=a_E'/a_E$ is the Einstein frame Hubble factor in $u$
time. Consequently \be H_E' \le 0 \ee and so it is impossible for
the Einstein frame scale factor to undergo a regular bounce. This is a null version of
the theorem for spatially flat FRW space-times that states \be
\dot{H}=-\frac{M_4^2}{2}(p+\rho) \le 0. \ee However, whilst the
latter is only a statement about Einstein's field equations, the
former is a statement about the string RG equations to all orders in
$\alpha'$. In fact there is a subtlety here, there are solutions in which $a$ reverses from expansion to contraction such as $a(U)=U$ or more generally when $a_s \rightarrow U+\alpha U^3$ as $U \rightarrow 0$. We evade the theorem because $a(U)$ passes through zero where $H$ is infinite. If the $x_i$ are non-compact then these are completely regular solutions, and their existence arises because in this case there is not a physical expansion or contraction associates with the change in $a_E(U)$, in fact we may change coordinates to a frame in which the new scale factor is expanding when $a_E$ is contracting. If the $x_i$ are compact then these solutions are generalized orbifolds of the type considered in \cite{Seiberg}. It is known from the case considered in \cite{Seiberg} that divergences arise in the string S-matrix. These were further connected to a gravitational instability in \cite{HorowitzPolchinski}. We may note that these higher dimensional orbifolds also satisfy the criteria discussed for the formation of black holes suggesting that the perturbative S-matrix will also be ill-defined in this higher dimensional setup.

\end{document}